\documentclass[
    ,final            %
  ]
  {aipproc}

\layoutstyle{6x9}

\usepackage{graphicx}%
\usepackage{dcolumn}%
\usepackage{bm}%
\usepackage{amsmath}
\usepackage{amsfonts}
\usepackage{subfigure} %

\def\ls{\mathrel{\lower4pt\vbox{\lineskip=0pt\baselineskip=0pt
           \hbox{$<$}\hbox{$\sim$}}}}
\def\gs{\mathrel{\lower4pt\vbox{\lineskip=0pt\baselineskip=0pt
           \hbox{$>$}\hbox{$\sim$}}}}
\def\drawbox#1#2{\hrule height#2pt

\hbox{\vrule width#2pt height#1pt \kern#1pt
              \vrule width#2pt}
              \hrule height#2pt}

\def\Asym#1#2{\vcenter{\vbox{\drawbox{#1}{#2}
              \kern-#2pt       %
              \drawbox{#1}{#2}}}}

\begin{document}

\title{$U(1)_{B-L}$ Sneutrino Dark Matter Detection with the IceCube Neutrino Telescope}


\author{Katherine Richardson-McDaniel}{
  address={Department of Physics \& Astronomy, University of New Mexico, Albuquerque, NM 87131, USA}
}

\begin{abstract}
We investigate the prospects for indirect detection of right-handed sneutrino dark matter at the IceCube neutrino telescope in a $U(1)_{B-L}$ extension of the MSSM. The capture and annihilation of sneutrinos inside the Sun reach equilibrium, and the flux of produced neutrinos is governed by the sneutrino-proton elastic scattering cross section, which has an upper bound of $8 \times 10^{-9}$ pb from the $Z^{\prime}$ mass limits in the $B-L$ model. Despite the absence of any spin-dependent contribution, the muon event rates predicted by this model can be detected at IceCube since sneutrinos mainly annihilate into leptonic final states by virtue of the fermion $B-L$ charges. These subsequently decay to neutrinos with 100\% efficiency. The Earth muon event rates are too small to be detected for the standard halo model irrespective of an enhanced sneutrino annihilation cross section that can explain the recent PAMELA data. For modified velocity distributions, the Earth muon events
  increase substantially and can be greater than the IceCube detection threshold of 12 events $\mathrm{km}^{-2}$ $\mathrm{yr}^{-1}$. However, this only leads to a mild increase of about 30\% for the Sun muon events. The number of muon events from the Sun can be as large as roughly 100 events $\mathrm{km}^{-2}$ $\mathrm{yr}^{-1}$ for this model.
\end{abstract}

\maketitle

\section{Introduction}

Supersymmetry is a front-runner candidate to address the hierarchy problem of the standard model (SM), and it has a natural dark matter candidate, namely the lightest supersymmetric particle (LSP), which can have the correct thermal relic abundance. A minimal extension of the SM gauge group, motivated by the nonzero neutrino masses, includes a gauged $U(1)_{B-L}$ gauge symmetry~\cite{mohapatra} ($B$ and $L$ are baryon and lepton number respectively). Anomaly cancellation then implies the existence of three right-handed (RH) neutrinos and allows us to write the Dirac and Majorana mass terms for the neutrinos to explain the light neutrino masses and mixings.

The $B-L$ extended MSSM also provides new dark matter candidates: the lightest neutralino in the $B-L$ sector~\cite{khalil,ADRS} and the lightest RH sneutrino~\cite{inflation2}.
In this work we will focus on the sneutrino as the dark matter candidate\footnote{It is also possible to have successful inflation in the context of the $U(1)_{B-L}$ model~\cite{inflation1}. In this case the dark matter candidate (the RH sneutrino) can become a part of the inflaton field and thereby gives rise to a unified picture of dark matter, inflation and the origin of neutrino masses~\cite{inflation2}.}. The candidate is made stable by invoking a discrete $R$-parity, but in the context of a $B-L$ symmetry, a discrete matter parity can arise once the $U(1)_{B-L}$ is spontaneously broken~\cite{rparity}. The $B-L$ gauge interactions can yield the correct relic abundance of sneutrinos if the $U(1)_{B-L}$ is broken around the TeV scale. Recently, it has been shown that it is possible to explain the positron excess observed in the PAMELA data~\cite{pameladata} in the context of a low scale $B-L$ extension of the MSSM~\cite{ADRS,Allahverdi:2009ae,Dutta:2009uf}.

The RH sneutrino of this $B-L$ extended model can be detected when it elastically scatters off a nucleus. The sneutrino-proton scattering cross section is in the $10^{-11}-10^{-8}$ pb range from the $Z^{\prime}$ mass limits \cite{tevatron, LEP}, large enough to be probed in the ongoing and upcoming dark matter direct detection experiments ~\cite{inflation2, Allahverdi:2009se}. In addition, annihilation of sneutrinos at the present time produces LH neutrinos. It is interesting to investigate the possibility of indirect detection of sneutrino dark matter by using final state neutrinos in the IceCube neutrino telescope. This ongoing experiment plans to probe the neutrino flux arising from the annihilation of gravitationally trapped dark matter particles in the Sun and the Earth.
\section{The $U(1)_{B-L}$ Model and Neutrino Production}

Since this $B-L$ is a local gauge symmetry, we have a new gauge boson $Z^{\prime}$ (and its supersymmetric partner). In the minimal model, we also have two new Higgs fields $H^{\prime}_1$ and $H^{\prime}_2$ (that are SM singlets) and their supersymmetric partners. The vacuum expectation values (VEVs) of these Higgs fields break the $B-L$ symmetry~\cite{Allahverdi:2009se} and result in a mass $m_{Z^{\prime}}$ for the $Z^{\prime}$ gauge boson. Various $B-L$ charge assignments are allowed by anomaly cancelation. We choose the charge assignment in \cite{Allahverdi:2009se}, where $H^{\prime}_2$ couples to the RH neutrinos and gives rise to a Majorana mass upon spontaneous breakdown of the $U(1)_{B-L}$. Choosing these Majorana masses in the $100~{\rm GeV}-1$ TeV range, we have three (dominantly RH) heavy neutrinos and three (dominantly LH) light neutrinos. The masses of the light neutrinos are obtained via the see-saw mechanism.

A natural dark matter candidate in this model is the lightest sneutrino ${\widetilde N}$. We consider the following two cases:
\begin{itemize}
{\item
{\bf Case 1}: A generic case where a solution to the positron excess observed by PAMELA is not sought. In this case the dominant annihilation channels are the $S$-wave processes ${\widetilde N} {\widetilde N} \rightarrow N N$ and ${\widetilde N}^* {\widetilde N}^* \rightarrow N^* N^*$ via $t$-channel exchange of ${\widetilde Z}^{\prime}$.
There are also ${\widetilde N} {\widetilde N}^* \rightarrow N N^*,~f {\bar f}$ annihilation modes via $s$-channel exchange of a $Z^{\prime}$ or $B-L$ Higgs fields, but these are $P$-wave suppressed and can be completely neglected (particularly at the present time). In this case the annihilation cross-section has the nominal value $\sim 3\times 10^{-26}$ cm$^3$/sec (dictated by thermal freeze out) at all times. The RH neutrinos produced from dark matter annihilation quickly decay to LH neutrinos and the MSSM Higgs.} Assuming that the mass difference between the RH sneutrinos and RH neutrinos is small, the RH neutrinos are produced non-relativistically, and hence each LH neutrino and Higgs receives an energy equal to half of the sneutrino mass. This produces a delta function in the energy of the LH neutrinos at one-half the mass of the sneutrino dark matter.  A small portion of muon neutrinos from this initial annihilation state are scattered via neutral current interactions inside the Sun to lower energies. This produces a slight bump in the neutrino spectrum at low energies. As expected for monochromatic neutrinos, the spectrum of muons from charge current interactions in {\bf{case 1}} has a linear dependence on energy.
{\item
{\bf Case 2}: In this case the PAMELA puzzle is addressed via Sommerfeld enhancement of sneutrino annihilation at the present time~\cite{Allahverdi:2009ae}. The dominant annihilation channel is ${\widetilde N}^* {\widetilde N} \rightarrow \phi \phi$ via the $s$-channel exchange of the new scalar Higgs fields, the $t$ or $u$-channel exchange of a ${\widetilde N}$, and the contact term $\vert {\widetilde N} \vert^2 \phi^2$. The cross section for annihilation to the $\phi \phi$ final state at the present time is required to be $3 \times 10^{-23}$ cm$^3$/sec in order to explain the PAMELA data. Sufficient Sommerfeld enhancement is obtained as a result of the attractive force between sneutrinos due to the $\phi$ exchange provided that the mass of $\phi$ is small ($<20$ GeV). The $\phi$ subsequently decays into fermion-antifermion pairs very quickly via a one-loop diagram, and it mostly produces $\tau^+\tau^-$ final states by virtue of the fermion $B-L$ charges~\cite{Allahverdi:2009ae}.} RH neutrinos constitute about 10\% of the annihilation final states, while $\phi\phi$ compose the remaining $90\%$ of the branching fraction. This branching fraction is necessary to provide a high enough leptonic particle rate to fit the PAMELA data. For $4\,{\rm GeV}<m_\phi<20$ GeV, the final states are mostly taus (74\%) and b quarks (16\%), where the dominance of tau final states is a result of the fermion $B-L$ charges. The LH neutrinos in this case arise from the three-body decay of taus and bottom quarks and are spread in energy signal. The delta function from the neutrino channel at the detector is subdominant to the other annihilation channels. The spectrum of muons is softened compared to {\bf{case 1}}.

\end{itemize}
Both the {\bf{case 1}} and {\bf{case 2}} scenarios of our model display a crucial signature difference when compared to the standard neutralino LSP in the MSSM. The energy distribution of the produced LH neutrinos from the RH neutrino decay is a delta function occurring at half of the sneutrino mass. Other annihilation channels in this model, as well as those available in the MSSM, produce additional neutrino signal via three-body decays such as  $\tau^- \rightarrow e^- \nu_{\tau}\bar{\nu_{e}}$. This difference opens up a significant possibility to differentiate between the $B-L$ model and the MSSM with the help of the differential energy spectrum of the detector event rates.

\section{IceCube Neutrino Signal Results} \label{Neutrino Flux}

Sneutrino annihilation in the Sun and the Earth produces an expected neutrino flux and charge current interactions in IceCube. The muon track signal can be differentiated from the cosmic ray induced background by selecting upward-going and contained muon events, subtracting atmospheric neutrino background expectations and making angle cuts on the muon tracks. The muon and neutrino fluxes are modeled by calculating the number of gravitationally captured sneutrinos and then considering the propagation and detection of the produced neutrinos. Using DarkSUSY and WimpSim tables, our calculations account for neutrino oscillation, loss via charged current interactions and scattering via neutral current interactions~\cite{DarkSUSY, WimpSim}. DarkSUSY default parameters are used. For both {\bf{case 1}} and {\bf{case 2}}, the maximum spin-independent cross section allowed by the $Z^{\prime}$ limits is used. We assume equal branching to the three flavors of LH neutrinos, since the results of~\cite{Allahverdi:2009se} do not depend critically on the choice of neutrino flavor branching ratios in either case.

\begin{figure}[t]
\centering
\includegraphics[height=.2\textheight]{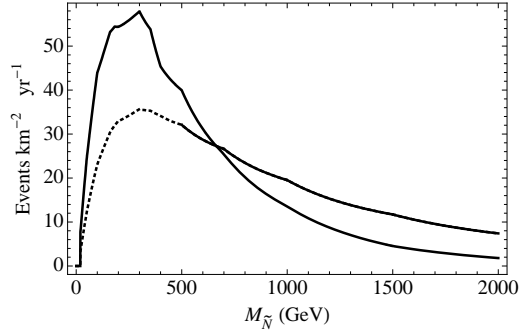}
\centering
\caption{\label{fig:MassPlotsSunMuons}Total muon rates detected at the Earth from annihilation of sneutrino dark matter in the Sun as a function of the sneutrino mass. The results are for one year of detection with IceCube. {\bf{Case 1}} ({\bf{case 2}}) is the highest (lowest) peaked line. The dotted line denotes the mass range where one cannot explain the PAMELA data using {\bf{case 2}} anymore.}
\end{figure}

Fig.~\ref{fig:MassPlotsSunMuons} shows our results for the total muon rate integrated over energy as a function of
the sneutrino mass $m_{\widetilde N}$ for annihilation in the Sun\footnote{The apparent discrete nature of these plots occurs because only a few values of sneutrino mass are recorded in the WimpSim tables used by DarkSUSY; the program interpolates between these points. The effect is numerical and not physical.}.

The general decrease of the event rates for higher $m_{\widetilde N}$ is reflective of the decrease of the neutrino flux due to the kinematic suppression of sneutrino capture ($\sim 1/m_{\widetilde N}$ for large masses). The linear increase at low $m_{\widetilde N}$ is explained by the linear dependence of the cross section for charged current interactions on the energy of neutrinos at the detector ($\sim m_{\widetilde N}$). LH neutrinos are produced in two-body decays in {\bf {case 1}} versus three-body decays in {\bf {case 2}}, and hence have a higher energy. Thus at lower values of sneutrino mass, the cross section for conversion of neutrinos to muons at the detector is larger in {\bf {case 1}}. However, for large sneutrino masses {\bf {case 1}} has a smaller signal than {\bf 
 {case 2}} since neutrinos get absorbed via charged current interactions or lose energy via neutral current interactions inside the Sun more efficiently because of their larger energy.

According to refs.~\cite{C.DeClercq,Abbasi:2009uz}, in the case of the Earth more than 12 events are needed for a DM mass between 70 GeV and 4 TeV. In the case of the Sun the number of events needed drops linearly as a function of mass starting from 300 events at 70 GeV down to 70 events at 300 GeV. Beyond 300 GeV up to 4 TeV, the number of events needed remains fixed at 70. This provides a hint that one could detect the event rates caused by sneutrinos despite some differences between the sneutrino and neutralino dark matter spectra used to calculate these sensitivities. Hence, it might be possible to detect muon neutrinos produced by sneutrino annihilation for sneutrino masses around 300 GeV for the Sun, cf.~Fig.~\ref{fig:MassPlotsSunMuons}. Note that a large range of masses would be accessible with only an order of magnitude improvement in sensitivity.

For a known sneutrino mass, observation of a muon signal exceeding the number given in Fig.~\ref{fig:MassPlotsSunMuons} will rule out the $B-L$ model. The largest number of muon events from the Sun in the entire depicted mass range is 58 $\mathrm{km}^{-2}$ $\mathrm{yr}^{-1}$ (36 $\mathrm{km}^{-2}$ $\mathrm{yr}^{-1}$) for {\bf {case 1}} ({\bf {case 2}}).

\section{Dark Matter Disc in the Milky Way} \label{DM Disc in the Milky Way}

In our analysis, we assumed a Gaussian-like velocity distribution for dark matter particles
with a three dimensional velocity dispersion of $\sigma_v = 270$ $\mathrm{km}$ $\mathrm{sec}^{-1}$ and $\vert v_{Sun}\vert=220$ $\mathrm{km}$ $\mathrm{sec}^{-1}$ for the velocity of the solar system with respect to the halo. However, there are recent speculations about the existence of a dark matter thick disc in the Milky Way in addition to the baryonic one, see e.~g.~\cite{Read:2008fh,Read:2009iv}. This dark matter disc would have different ranges for the solar system velocity and velocity dispersion: $\vert v_{Sun}\vert\approx 0-150$ $\mathrm{km}$ $\mathrm{sec}^{-1}$ and $\sigma_v \approx 87-1
 56$ $\mathrm{km}$ $\mathrm{sec}^{-1}$~\cite{Bruch:2009rp}. A modified velocity distribution substantially enhances the Earth muon rate for the sneutrino dark matter beyond the detection threshold of 12 $\mathrm{km}^{-2}$ $\mathrm{yr}^{-1}$ over a large part of the velocity parameter space~\cite{Allahverdi:2009se}. It also raises the maximum Sun muon rate to 78 events $\mathrm{km}^{-2}$ $\mathrm{yr}^{-1}$ (48  $\mathrm{km}^{-2}$ $\mathrm{yr}^{-1}$) in {\bf {case 1}} ({\bf {case 2}}).
\section{Comparison with mSUGRA} \label{Comparison}

Minimal supergravity (mSUGRA) is a constrained version of the MSSM with the lightest neutralino as its dark matter candidate. In this section we compare the $B-L$ model with the mSUGRA hyperbolic branch/focus point mSUGRA scenarios, where the dark matter has a large Higgsino component and $m_0$ is very large. The neutralino has a large capture rate in this region due to a large Higgsino component that results in a large spin-dependent scattering cross section via $Z$ exchange. 

\begin{figure}[h]
\centering
\includegraphics[height=.2\textheight]{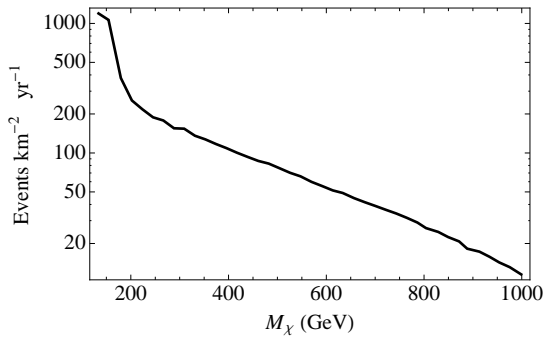}
\caption{\label{fig:MassPlotSunRateFocusPoints}Total Sun-annihilation muon rates inside the detector for mSUGRA hyperbolic branch/focus point scenarios as a function of the neutralino mass.
The results are for one year of detection with IceCube.}
\end{figure}

Fig.~\ref{fig:MassPlotSunRateFocusPoints} shows the total Sun muon rate as a function of the neutralino mass for mSUGRA hyperbolic branch/focus points. A comparison with Fig.~\ref{fig:MassPlotsSunMuons} shows that these scenarios always have a higher total muon rate in the plotted mass range than the $B-L$ model. The hyperbolic branch/focus point models yield larger muon rates by between more than one order of magnitude and a factor of $1.5$ for dark matter masses in the $100-800$ GeV range. These higher rates are explained by the spin-dependent scattering cross sections, which are a few orders of magnitude larger than the limits on the spin-independent cross section for the sneutrino dark matter. The spin-dependent scattering cross section for the $B-L$ model is zero because $U(1)_{B-L}$ is a vectorial symmetry. Since the Sun mainly consists of hydrogen, the spin-dependent piece contributes dominantly for the mSUGRA case.

However, it is interesting that despite having a much smaller scattering cross section, the $B-L$ model can yield muon rates that are roughly comparable to the mSUGRA scenarios. Sneutrino annihilation dominantly produces leptons, i.e., RH neutrinos in {\bf {case 1}} and taus in {\bf {case 2}}, which subsequently decay to LH neutrinos $100\%$. On the other hand, neutralino annihilation in the hyperbolic branch/focus point scenarios dominantly produces quark final states that have a small branching ratio for decay to neutrinos. Furthermore, despite lower event rates, sneutrino dark matter still produces a distinctive linear spectrum in the muon flux. This feature is caused by the delta function in energy for the neutrino spectrum and can be used to distinguish between the $B-L$ model and the hyperbolic branch/focus point scenarios as long as energy binning of the differential muon rate with respect to the energy is precise enough at IceCube.

\begin{theacknowledgments}
  The author wishes to thank Spencer Klein and Carsten Rott for valuable discussions. This presentation is based on work~\cite{Allahverdi:2009se} with Rouzbeh Allahverdi, Sascha Bornhauser and Bhaskar Dutta.
\end{theacknowledgments}

\bibliographystyle{aipproc}   %

\begin{thebibliography}{99}

%
%
%


%
%


%
%


\bibitem{mohapatra}
%
R.N.~Mohapatra and R.E.~Marshak,
  %
  %
Phys.\ Rev.\ Lett.\  {\bf 44}, 1316 (1980)
[Erratum-ibid.\  {\bf 44}, 1643 (1980)].
  %


\bibitem{khalil}
S.~Khalil and H.~Okada,
  %
arXiv:0810.4573 [hep-ph];
  %
S.~Khalil and A.~Masiero,
  %
  Phys.\ Lett.\  B {\bf 665}, 374 (2008).


\bibitem{ADRS}
  R.~Allahverdi, B.~Dutta, K.~Richardson-McDaniel and Y.~Santoso,
  %
  %
  Phys.\ Rev.\  D {\bf 79}, 075005 (2009).
  %
  %


\bibitem{inflation2}
  R.~Allahverdi, B.~Dutta and A.~Mazumdar,
  %
  Phys.\ Rev.\ Lett.\  {\bf 99}, 261301 (2007).
 %
  %
%


\bibitem{inflation1}
 R.~Allahverdi, A.~Kusenko and A.~Mazumdar,
  %
  JCAP {\bf 0707}, 018 (2007).
  %
  %


\bibitem{rparity}
  S.~P.~Martin,
  %
  Phys.\ Rev.\  D {\bf 54}, 2340 (1996)
  [arXiv:hep-ph/9602349].
  %


\bibitem{pameladata}
O. Adriani, {\it et al.}, arXiv:0810.4995; arXiv:0810.4994.
%


\bibitem{Allahverdi:2009ae}
  R.~Allahverdi, B.~Dutta, K.~Richardson-McDaniel and Y.~Santoso,
  %
  arXiv:0902.3463 [hep-ph] (to appear in Phys. Lett.B).
  %

\bibitem{Allahverdi:2009se}
  R.~Allahverdi, S.~Bornhauser, B.~Dutta and K.~Richardson-McDaniel,
  %
  Phys.\ Rev.\  D {\bf 80} (2009) 055026
  [arXiv:0907.1486 [hep-ph]].
  %

\bibitem{Dutta:2009uf}
  B.~Dutta, L.~Leblond and K.~Sinha,
  %
  %
  arXiv:0904.3773 [hep-ph].
  %


%
%


%
%


\bibitem{tevatron}
  T.~Aaltonen {\it et al.}  [CDF Collaboration],
  %
  %
  Phys.\ Rev.\ Lett.\  {\bf 99} (2007) 171802
  [arXiv:0707.2524 [hep-ex]].
  %



\bibitem{LEP}
  M.~S.~Carena, A.~Daleo, B.~A.~Dobrescu and T.~M.~P.~Tait,
  %
  Phys.\ Rev.\  D {\bf 70} (2004) 093009
  [arXiv:hep-ph/0408098].
  %

%
%
%
%
%
%
  %

%
%

 %
\bibitem{C.DeClercq}
    C. De Clercq for the IceCube Collaboration,
    ``Search for Dark Matter with the AMANDA and IceCube Neutrino Detectors'',
    Presented at the Identification of Dark Matter 2008, Stockholm, Sweden, 18-22 August 2008; Proceedings of Science PoS (idm2008) 034.

%
\bibitem{Abbasi:2009uz}
  R.~Abbasi {\it et al.}  [ICECUBE Collaboration],
  %
  %
  Phys.\ Rev.\ Lett.\  {\bf 102} (2009) 201302
  [arXiv:0902.2460].
  %


%
\bibitem{DarkSUSY}
  P.~Gondolo, J.~Edsjo, P.~Ullio, L.~Bergstrom, M.~Schelke and E.~A.~Baltz,
  %
  JCAP {\bf 0407} (2004) 008
  [arXiv:astro-ph/0406204].
  %
%
\bibitem{WimpSim}
  M.~Blennow, J.~Edsjo and T.~Ohlsson,
  %
  JCAP {\bf 0801} (2008) 021
  [arXiv:0709.3898 [hep-ph]].
  %

%
%


%
\bibitem{Read:2008fh}
  J.~I.~Read, G.~Lake, O.~Agertz and V.~P.~Debattista,
  %
  arXiv:0803.2714 [astro-ph].
  %


%
\bibitem{Read:2009iv}
  J.~I.~Read, L.~Mayer, A.~M.~Brooks, F.~Governato and G.~Lake,
  %
  %
  arXiv:0902.0009 [astro-ph.GA].
  %


%
\bibitem{Bruch:2009rp}
  T.~Bruch, A.~H.~G.~Peter, J.~Read, L.~Baudis and G.~Lake,
  %
  arXiv:0902.4001 [astro-ph.HE].
  %


%
%
%
%
%
%
%


%
%
%
%
%
%
%
%
%
%
%
%
%
%
%
%
%
%
%
%
%
%
%
%
%
%
%
%
%
%
%
%
%
%
%
%
%
%
%
%
%
%
%
%
%

\end{thebibliography}

\end{document}